\documentclass[preprint,double-spaced,aps,apssymb]{revtex4}
\usepackage{amsmath,amssymb}
\usepackage{graphicx}
\usepackage{dcolumn}
\usepackage{bm}
\usepackage{subfigure}
\newcommand{\ber}{\begin{eqnarray}}
\newcommand{\eer}{\end{eqnarray}}
\newcommand{\bea}{\begin{equation}}
\newcommand{\eea}{\end{equation}}

\begin{document}

\title{Volume exclusion and elasticity driven directional transport: an alternative model for bacterium motility}

\author{A. Bhattacharyay}
\email{a.bhattacharyay@iiserpune.ac.in}
 \affiliation{Indian Institute of Science Education and Research, Pune, India}

\date{\today}

\begin{abstract}
On the basis of a model we capture the role of strong attractive interaction in suppressing the rotational degrees of freedom of the system and volume exclusion in keeping microscopic symmetry-breaking intact to result in super-diffusive transport of small systems in a thermal atmosphere over a large time scale. Our results, characterize such systems on the basis of having a super-diffusive intermediate regime in between a very small and large time scales of diffusive regimes. Although, the Brownian ratchet model fails to account for the origin of motility in actin polymerization propelled directional motion of bacterium like Listeria Monocytogene (LM) and similar bio-mimetic systems due to the presence of strong attractive forces, our model can account for the origin of directional transport in such systems on the basis of the same interactions.         
\end{abstract}
\maketitle

It is widely believed that the basis of driven directional motion of various motor proteins, bacterium etc., as observed in living systems or in similar environments, is the Brownian Ratchet (BR) mechanism \cite{ast,rei,han}. The BR mechanism essentially requires the assistance of noise, a drive which can be periodic in time and directionless over space and a potential which must have broken reflection symmetry (polar) at a small (microscopic) scale \cite{ast}. All of these ingredients are pretty much available in the biological world. The biological systems normally perform at standard room temperatures - so, thermal noise is present. Often the tubular tracks on which the motors move are made up of polar units and actin monomers are also polar in nature by polymerizing which some bacteria move. The motion of motor proteins and other bio-systems are invariably driven i.e. are associated with intermittent energy supply to drive them out of equilibrium as a prerequisite of having directional motion in any system. There have been attempts to model the motion of the bacterium Listeria Monocytogene \cite{the} and similar biomimetic systems \cite{arp,jin}, which moves by polymerizing an actin tail behind it, following BR paradigm \cite{pes,mog1,mog2}. But, subsequently found experimental facts go very much against the basic requirements of the BR theory of such systems. It has been found that the bacterium remains bound to its tail with a strong elastic force (40 pN approx.) during the course of its motion. The bacterium practically undergoes 20 times less Brownian motion than similar objects in the same environment because of this strong binding \cite{kuo}. These observations practically rule out the scope of BR mechanism to explain the origin of directional transport in such actin polymerization based motile systems like LM.  
\par
In the present paper we put forward a model based on a structured object unlike the BR models where the moving entity is a point particle. We show that the strong force of attraction between the parts of the system plays an important role in making the motion directional over a considerable time scale by the suppression of rotational degrees of freedom. The other most important observation, on the basis of the present model, is the role of volume exclusion. Volume exclusion helps the system retain the locally broken symmetry. This phenomenon we illustrate by an exactly solvable model in one dimension (1D). Then we extend the analysis to 2D and numerically show that the importance of the volume exclusion remains unaltered in keeping the locally broken symmetry being manifested globally. In experiments it has been revealed that the width of the actin tail, the bacterium polymerizes, is most important to keep the directionality of the motion. The reason for that clearly follows from the model we are presenting. Thus, the model we put forward here presents an alternative mechanism to the BR which can produce directional motion of driven microscopic systems in a noisy environment. 
\par
The most important understanding, in the present context, is the role of volume exclusion in controlling directional transport of a driven system. This is not possible to explore with a BR type model where the moving system is structureless (particle) and the very naturally occuring process of volume exclusion remains trivially taken care of in such model systems. Within the scope of our model, by virtue of having structures, we investigated the role of volume exclusion and have conclusively shown that turning it on or off has dramatic consequences on the transport properties of the system. This is an important basic understanding and might find immense applicability in engineering micro machines which could show bacterium like motility. On another important account our model differs from the BR models is there is no large scale global symmetry breaking by the repeatition of polar elements as we consider it here. Rather, the present model investigates the microscopic conditions under which the symmetry broken at smaller scale reflects at a larger scale.
\par
The important general result that we get on the basis of the present model, which mimics a LM type structured object, is strong attractive force within a properly structured object which obeys volume exclusion can make it move super-diffusively over a large time scale. If one considers time scales bigger than this range one eventually observes the system to behave diffusively with an enhanced diffusivity. There also are smaller scales of time than this intermediate super-diffusive regime where the system shows diffusive motion. These smaller time scale diffusion is an artifact of the system having structures. The different parts of the system can undergo thermal fluctuations before experiencing appreciable restoring stress reflects on this smallest time scale. In the intermediate time scales, over which the motion is super-diffusive, a correlation is maintained between the orientations of the moving system which does not very often change the direction of motion due to internal resistance to rotational degrees of freedom. We argue that the motile objects like LM directionally operate in this intermediate temporal regime. The length scales which are of the order of magnitude of cell sizes are compatible with this range of time scales and diffusive motion at a larger time may be even helpful to such systems in exploring the whole volume inside a cell. The content of the paper is arranged in the following way. In section I we put forward the exact analytic calculation for the 1D version of the model. Section II uses the knowledge of section I to extend the model to 2D and explains the numerical details. In section III we present the results of the numerical simulation of the model in 2D. We conclude with a discussion of the present results and comparisons of those with the known facts of LM motions in section IV.         
\section {1D model}
In 1D, let us consider a model system consisting of two particles connected by a spring. This system when subjected to thermal (white) noise at a particular temperature would come to equilibrium and eventually would never show any directional motion. To drive the system slightly out of equilibrium let us consider that one of the particles has its instantaneous velocity fed back to it just to disturb the fluctuation-dissipation relation that ensures the equilibrium. Application of this velocity feed back is a very general way of driving the system out of equilibrium which in practice can play the role of a variety of forcing that drives the system out of equilibrium. So, with the velocity feed back, the model will look like 
\ber
\dot{x_1} &=& -\alpha(x_1-x_2) +\beta\dot{x_1}+\eta_1\\
\dot{x_2} &=& -\alpha(x_2-x_1)+\eta_2.
\eer  
Here, $\alpha$ is the force constant of the spring, $\beta$ is the feed back coefficient i.e. a measure of the strength of the velocity feedback. The white noises $\eta_i$s have an average ($<\eta_i>$) equal to zero and the second moment $<\eta_i(t_1)\eta_j(t_2)>=2T\delta_{ij}\delta(t_1-t_2)$, where the Boltzmann-constant ($k_B$) has been absorbed by fixing the unit of temperature. Its interesting to note that at $\beta=0$ the second moment of the noise can be expressed in terms of $\alpha$ by adjusting the temperature to explicitly manifest the fluctuation-dissipation relation for the system. A nonzero $\beta $ breaks this relationship and makes the system go out of equilibrium. The presence of the nonzero beta serves another purpose for us namely breaking the symmetry. For the system to show directional motion in 1D there are two directions for it to move in, to make it take up one of them, the symmetry has to be broken which is achieved by the feed back as well. Now, the symmetry that has been broken at a microscopic scale i.e. at the scale of the object which would supposedly move, we want to see if this broken symmetry indeed shows up at global scales at which we expect to measure the directional motion of this model object. In what follows we will always keep $\beta$ small, in fact $\beta < 1$ keeping in mind that we are in the over damped regime of the dynamics. Consideration of a large $\beta$ would require inertial term to be taken into account which we do not want to consider in the present context.   
\par
Let us change the coordinates of the system and go to the internal coordinates $Z=x_1-x_2$ and the center of mass (CM) coordinate $X=(x_1-x_2)/2$. The model would be rewritten as 
\ber
\dot{Z}=-\frac{\alpha(2-\beta)}{1-\beta} Z +\xi_Z \\
\dot{X}= -\frac{\alpha\beta}{2(1-\beta)} Z + \xi_X
\eer 
In the above equations, the averages of the noise terms $\xi_i$ are clearly zero and there second moments are $<\xi_Z(t_1)\xi_Z(t_2)>=2T(1+\frac{1}{(1-\beta)^2})\delta(t_1-t_2)$ and $<\xi_X(t_1)\xi_X(t_2)>=\frac{T}{2}(1+\frac{1}{(1-\beta)^2})\delta(t_1-t_2)$. Eq.3 represents an equilibrium dynamics and we can readily get the probability distribution of the internal coordinate as $P_{eq}(Z)\propto \exp{(-\frac{\alpha(2-\beta)}{2TD_1(1-\beta)}Z^2)}$ (where $D_1=(1+\frac{1}{(1-\beta)^2})$). This is a distribution of zero average but we can readily calculate a nonzero average value for the internal coordinate Z which can never become negative (for $x_1>x_2$) due to volume exclusion. This restricts the range of integration falling on the positive side of the $Z$-axis only. Thus, the average internal span $<Z>_{eq}=\sqrt{\frac{2TD_1(1-\beta)}{\pi\alpha(2-\beta)}}$ and that immediately gives us, from Eq.4, an expression for the average velocity of the CM of the system as
\bea
<\dot{X}>=-\frac{\alpha\beta}{2(1-\beta)}<Z>_{eq}=-\frac{\alpha\beta}{2(1-\beta)}\sqrt{\frac{2TD_1(1-\beta)}{\pi\alpha(2-\beta)}}.
\eea
The above expression clearly shows that with $\beta\neq 0$ the system would have a nonzero average velocity. Thus, the symmetry broken at the microscopic scale is being reflected at macroscopic scales due to volume exclusion and being driven out of equilibrium. 
\par
A particular advantage for the system being in 1D is that symmetry once broken would always remain and would not be statistically averaged out over time because the system does not have any rotational degrees of freedom. So, a straight forward extension of this model in 2D is bound to fail. One has to think of a differently structured object in 2D, which due to its very structural properties, can actually suppress the rotational degrees of freedom and thus help broken symmetries at smaller scales reflect on a larger scale. We would like to keep the basic ingredients of the model like strong attraction between parts, volume exclusion etc intact in the 2D model as well. In fact, relation 5 clearly shows that the average velocity is proportional to the strength of the attractive interaction and we recognize the attractive interaction as an important ingredient. Moreover, the experimental observation of the LM having strong attractive interaction between the head and the tail encourages us to think of a model in 2D which is structurally similar to LM and preserves the basic ingredients of our 1D model.      
\section{2D model}
Keeping in mind the head tail structure of LM, in 2D we propose a three body model here consisting of three particles (say). The tail of the system consists of two particles which are energetically favoured to remain at a distance (this distance is tunable) from each other. The third particle, which represents the head of the system, is strongly attracted  by a spring force towards the other two particles making the tail (see the schematic diagram in Fig.1). In the absence of any noise the head will stay at the middle of the tail but, in the presence of the noise there will be fluctuations from this position and the system would get a triangular shep on average. To drive the system out of equilibrium we again put an instantenious velocity feedback on the head of the system. With all these in place, the model looks like,
\ber\nonumber
\dot x_1 &=& -\frac{\alpha}{1-\beta}(x_1-x_2)-\frac{\alpha}{1-\beta}(x_1-x_3)
+\frac{\eta_1\cos{\theta_1}}{1-\beta}\\\nonumber
\dot y_1 &=& -\frac{\alpha}{1-\beta}(y_1-y_2)-\frac{\alpha}{1-\beta}(y_1-y_3)
+\frac{\eta_1\sin{\theta_1}}{1-\beta}\\\nonumber
\dot x_2 &=& \alpha(x_1-x_2)-(1-\frac{\gamma}{\sqrt{(x_2-x_3)^2+(y_2-y_3)^2}})(x_2-x_3)+\eta_2\cos{\theta_2}\\\nonumber
\dot y_2 &=& \alpha(y_1-y_2)-(1-\frac{\gamma}{\sqrt{(x_2-x_3)^2+(y_2-y_3)^2}})(y_2-y_3)+\eta_2\sin{\theta_2}\\\nonumber
\dot x_3 &=& \alpha(x_1-x_3)+(1-\frac{\gamma}{\sqrt{(x_2-x_3)^2+(y_2-y_3)^2}})(x_2-x_3)+\eta_3\cos{\theta_3}\\\nonumber
\dot y_3 &=& \alpha(y_1-y_3)+(1-\frac{\gamma}{\sqrt{(x_2-x_3)^2+(y_2-y_3)^2}})(y_2-y_3)+\eta_3\sin{\theta_3}\\.
\eer
In this 2D model, $(x_1,y_1)$ is the centre of the head, the spring constant of the attractive force between the head and the tail is $\alpha$, the feed back constant is $\beta$ and the equilibrium separation between the particles in the tail is $\gamma$. The noise $\eta_i$s have been applied in such a way that the thermal force remains isotropic and that is why we picked up $\theta_i$s randomly between the range $0-2\pi$. All the $\eta_i$s represent Gaussian white noise. The volume exclusion would be effectively taken by considering that none of the particles can cross the line joining the other two as if all the particles remain tightly tethered to the other two and do not allow any passage through them. Mathematically, this can be achieved by demanding the quantity $(x1-x2)(y1-y3)-(x1-x3)(y1-y2) $ does not change sign over the whole interval of the time evolution of the system. This is a non-holonomic constraint what makes it extremely difficult (if not impossible) to solve this problem analytically. That is why, we will take here the numerical route to investigate this model.
\par
To have the constraint in place while evolving the system numerically, we would consider every particle to undergo a reflection, when it tries to cross the line joining the other two and would find out the new positions of all the three particles (consistent with the constraint) from the momentum and energy conservation equations. If such a situation, where the constraint is violated, does not arise, there is no question of going into all these complications. In this process, the particles are considered of the same mass and moment of inertia of the two particle connected by a line is that of a dumbbell. Consider the situation when the head (particle 1) hits the line joining the particles on the tail (base line say) with a velocity $u=u_{\parallel}+u_{\perp}$ where $u_\parallel$ is the component of $u$ parallel to the base line and $u_\perp$ is the perpendicular component of $u$ to the same base line (from now on all the parallel and perpendicular notations would mean parallel and perpendicular to this line). After suffering reflection the velocity of the particle 1 becomes $v^\prime=v^\prime_\parallel+v^\prime_\perp$, the velocity of the center of mass of the tail is $v^{\prime\prime}=v^{\prime\prime}_\perp$, the angular velocity of the tail about its cm is $\omega$.  Let us also consider that the particle 1 has hit the tail at a distance $X$ ($X\leq R$) from its cm and the spread of the tail is $2R$. In such a situation, the momentum conservation gives us
\ber\nonumber
u_\parallel &=& v^\prime_\parallel \\\nonumber
u_\perp &=& -v^\prime_\perp + v^{\prime\prime}_\perp\\
u_\perp X &=& -v^\prime_\perp X + 2R^2\omega .
\eer   
The energy conservation results in the relation
\bea
|u|^2=|v^\prime|^2+2|v^{\prime\prime}_\perp|^2+2R^2\omega^2.
\eea
We can solve these four equations for the four unknowns namely $v^\prime_\perp$, $v^\prime_\parallel$, $v^{\prime\prime}_\perp$ and $\omega$ and that would give the corrections for the positions of the three particles when such a reflection happens to keep the volume exclusion constraint in place. Note that, consideration of this perfectly elastic collision is there to simplify the process of imposing the constraint, and we assert that, inelasticity to some extent would not be in conflict with the generality of result we will arrive at in the next section. However, the role of inelasticity can later be explored as a specific issue to understand one of the variants of the present model. Another point regarding the limitations of the present way of imposing the constraint is worth noting - in the simulations we are restricted to take not very large values of the $\alpha$ and $\beta$ and not very small value of $\gamma$ so as to keep $\omega$ sufficiently small. Otherwise, the tail will rotate so much that the relative orientations of the particles 2 and 3, which make the tail, with respect to particle 1 would change and that would again violate the constraint. Nevertheless, we will have a fairly large scale of the parameters mentioned above while following the present scheme of computations to establish the role of elasticity and volume exclusion in producing directional motion.
\par
In Fig.1 (schematic diagram of the model), when the constraint is in place, the thermal fluctuations would on average make the system take the shape of an isosceles triangle. Any, deviation of this shape would stretch sides in such a way that the resulting restoring force will try to bring the cm of tail to head vector $\vec{V}$ towards its previous orientation. The origin of this restoring force is the attraction between the head and the tail and one can easily qualitatively understand that the strength of the restoring force should increase with the force constant $\alpha$. This restoring force is responsible for the suppression of the rotational degrees of freedom of the system. But, when the volume exclusion is violated, the system undergoes such dramatic change in its internal structure that the $\vec{V}$ cannot be restored to its previous direction and the motion of the system should undergo drastic irreversible change in directions. Moreover, a wider tail i.e. a wider base of the triangular system should prevent abrupt change in the direction of motion of the system by simply making the volume exclusion apply more frequently and we guess at this point to register enhanced directionality in the motion of the system as the width of its tail increases. Thus, we qualitatively understand by looking at the model that a strong attractive interaction at least from two different directions between the head and the tail might go against the requirements of the BR model but are instrumental in the suppression of the rotational degrees of freedom of the system and facilitate the average directional motion.

\section{Results of the simulation}
The system does not show any directional motion when the constraint is not in place. Fig.2a shows the space covered by the system (cm has been plotted) over an interval of $10^7$ time steps each step size being 0.0001 in arbitrary units. The spring constant $\alpha =1.0$ and the noise strength $\eta_i=.001$, feedback constant $\beta=0.99$ and the width of the tail $\gamma = 0.1$. To have an estimate of how far the system moves compared to the system size one should use the tail width as a reference unit and that shows the system practically is undergoing structural fluctuations sitting at the same place. A dramatic change in the transport happens when we apply the volume exclusion keeping all other parameters the same. Fig.2b shows the trajectories of the system as it evolves over the same period of time with volume exclusion on. To show that the degree of directionality of the average motion of the system indeed increases with the width of the tail, in Fig.2c and 2d we have plotted the trajectories of the system when $\gamma$ is equal to 0.01 and 0.05 (all other parameters are the same and of course the volume exclusion is there) and we can notice the gradual increase in spread of the system with $\gamma$.   
\par
Suppression of the rotational degrees of freedom is clearly manifesting in the system having longer stretches of directional motion with lesser number of turn around. At a very large time scale, the motion would definitely be diffusive with an enhanced diffusivity, however, over a range of intermediate time scales the motion as appears from the nature of the trajectories is probably ballistic. This intermediate time scale is obviously selected by the structural details of the model system. In the case of the bacterium LM one can argue that the size of the cell which the bacterium invades being finite with respect to the size of the bacterium such an intermediate time scale of effective directional motion would be quite useful in avoiding frequent collisions with the cell boundaries. Moreover, the diffusive motion at a larger time scale would actually help LM in exploring the whole volume inside the cell for nutrition or whatever. Thus, the model reveals a clear dynamical motive of the structural properties of actin polymerization propelled bacterium.
\par
In Fig.3a and b we have plotted the directional correlation of the system without and with volume exclusion respectively. A vector has been drawn from the cm of the system at $t=t$ to $t=t+1000$ and the correlations between the angles that all such vectors make with the initial one have been numerically calculated for $10^6$ such vectors sequentially appearing over the time of evolution. Without feedback the correlation falls to zero after showing a peak in the negative direction. This negative peak is an artifact of the model. Since the head is strongly attracted by the tail it would on average pretty regularly change sides through the base line in the absence of the volume exclusion constraint and that gives rise to this peak at a characteristic time on the negative side. However, in Fig.3b we see a very long range correlation in the average directions of motion of the system and that explains the origin of the intermediate time scale at which the system follows an almost straight path. To understand the present scenario even more clearly, we have plotted (Fig.4) the log of root mean square distance $D$ of system from its position at $t=0$ against log of time $t$ for three different $\alpha$ values (1.0.2.0 and 3.0) when the volume exclusion is on and have compared that with the case when volume exclusion is absent ($\alpha = 1.0$). When volume exclusion is absent (call this graph 1) the model system shows three distinct regimes. The first part over a very short length scale is almost diffusive (exponent $\sim 0.45$). This part represents the internal fluctuations of the components of the system before it starts feeling appreciable restoring force. This part should be manifested irrespective of the volume exclusion being there or not and Fig.4 clearly demonstrates that. In the graph 1, the middle apparently flat part is due to the negative peak on the correlation plot which shows that there is a regular direction reversal, hence, enhanced suppression of the average motion. At even higher time scales the system eventually starts going diffusive may be due to some rare events that effectively gives it some translations. With the volume exclusion on, in this middle regime of time scales, the system shows very much super diffusive (almost ballistic) transport due to the long directed average stretches of the trajectories we have seen in Fig.2. As has been argued before, the transport of the system eventually becomes diffusive at very large time scales. Its interesting to note that with higher $\alpha$ i.e. stronger attractive force between the head and the tail the average directional motion of the system would be enhanced by a constant factor since the higher $\alpha$ graphs are lying above with practically the same slope.        

\section{Discussions}
The present model, being guided by the essential structural features of LM, reveals a fundamental role of elasticity and volume exclusion in enhancing directional transport of driven micro-systems. It puts forward an alternative scenario of the BR mechanism which cannot explain the LM motion. The essential role of strong attraction between the head and the tail of LM is understood as required for suppression of rotational degrees of freedom whereas the presence of this strong attraction entirely goes against the requirements of the BR model for such systems. The role of the width of the tail is also manifested in enhancing the directional transport. In fact, in actual experiments with LM like biomimetic systems people have seen that a not well formed thin tail does not help maintain the directionality and the system veers around more frequently. The transport properties of such systems are clearly characterized by three distinct time scales over which the motions are diffusive, super-diffusive and diffusive respectively. All the three regimes are essential for the LM motion to take place. The diffusive regime at the smallest time scale, which manifests internal fluctuations up to the limit the system starts feeling too much of restoring force, would help make room for the polymerization to take place. In the case of the bacterium LM this regime probably has to be tracked in the fluctuations of the membrane near the tail. The intermediate regime of the super diffusive transport is the one that makes the bacterium move directionally. This directional transport is enhanced to some extent by the stronger attractive interaction and the wider tail. It would be interesting to monitor if a bacterium adjusts the width of its tail when brought from a smaller enclosure to a larger one in real experiments. As the present theory proposes the dynamical motive of having such a structure is having the directional transport at relatively smaller time scales and diffusive transport at larger time scales, the bacterium might be equipped with some mechanism of having controls over these regimes. From this stand point, one can experimentally look at any adjustment in the construction of the tail by the bacterium to see if it could feel the presence of a wider/shorter region to move in. One can also experimentally evaluate the degree of directional transport with the variation of the tail width and can compare that with a suitable model which has the basic structure of the one presented here. The same investigation can also possibly be done if one can control the variation of the attractive force between the bacterium and its tail. Identification of the three regimes of motion in the case of LM or similar systems can conclusively prove the validity of the model. 
\par
In trying to understand the origin of directional transport for LM like systems, the role of elasticity and specially volume exclusion has emerged to be central. The volume exclusion is such a naturally occurring obvious fact that we often overlook its role. However, a clear understanding of the role of volume exclusion in the directional transport would prove extremely useful in designing small machines with ability to show directional transport in a thermal atmosphere. The principle ingredients of designing a class of such micro-motors, as par our present analysis, is a head attracted by a tail at least from two different directions and effective volume exclusion. We very much hope that understanding of the present mechanism of directed transport would be useful in direct applications and engineering of microscopic motile objects. \\  
{\bf Acknowledgement} I acknowledge very useful discussions with Prof. Amos Maritan.

\newpage

\begin{figure}
\begin{center}
\includegraphics[width=12cm,angle=0]{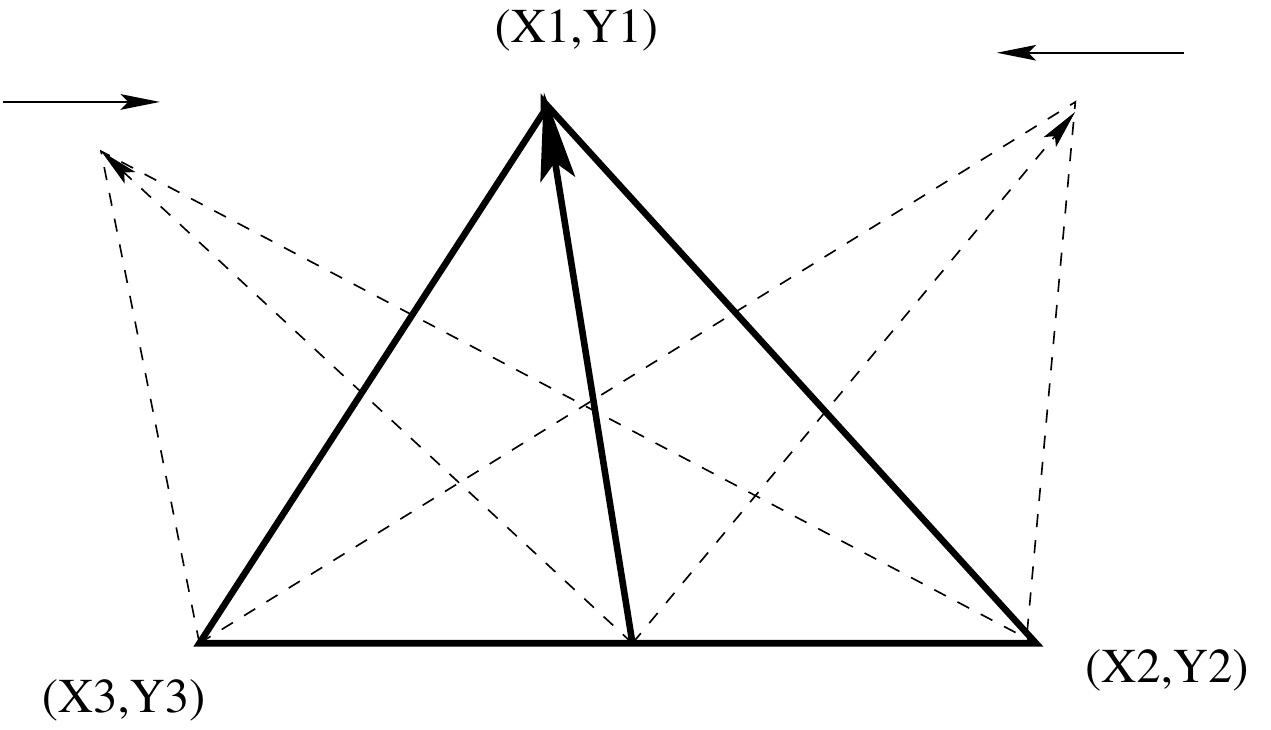}
\caption[Figure 1]{Schematic diagram of the model. Following deviations in the position of the head ($x_1,y_1$), on either sides of the central position, the direction of the resultant restoring force due to unequal streatching of the sides have been qualitatively represented by the arrows. The central vector, from the cm of the tail to the head is the {$\vec{V}$} which determines the direction of motion of the system.}
\end{center}
\end{figure}

\newpage

\begin{figure}
\begin{center}
\includegraphics[width=12cm,angle=0]{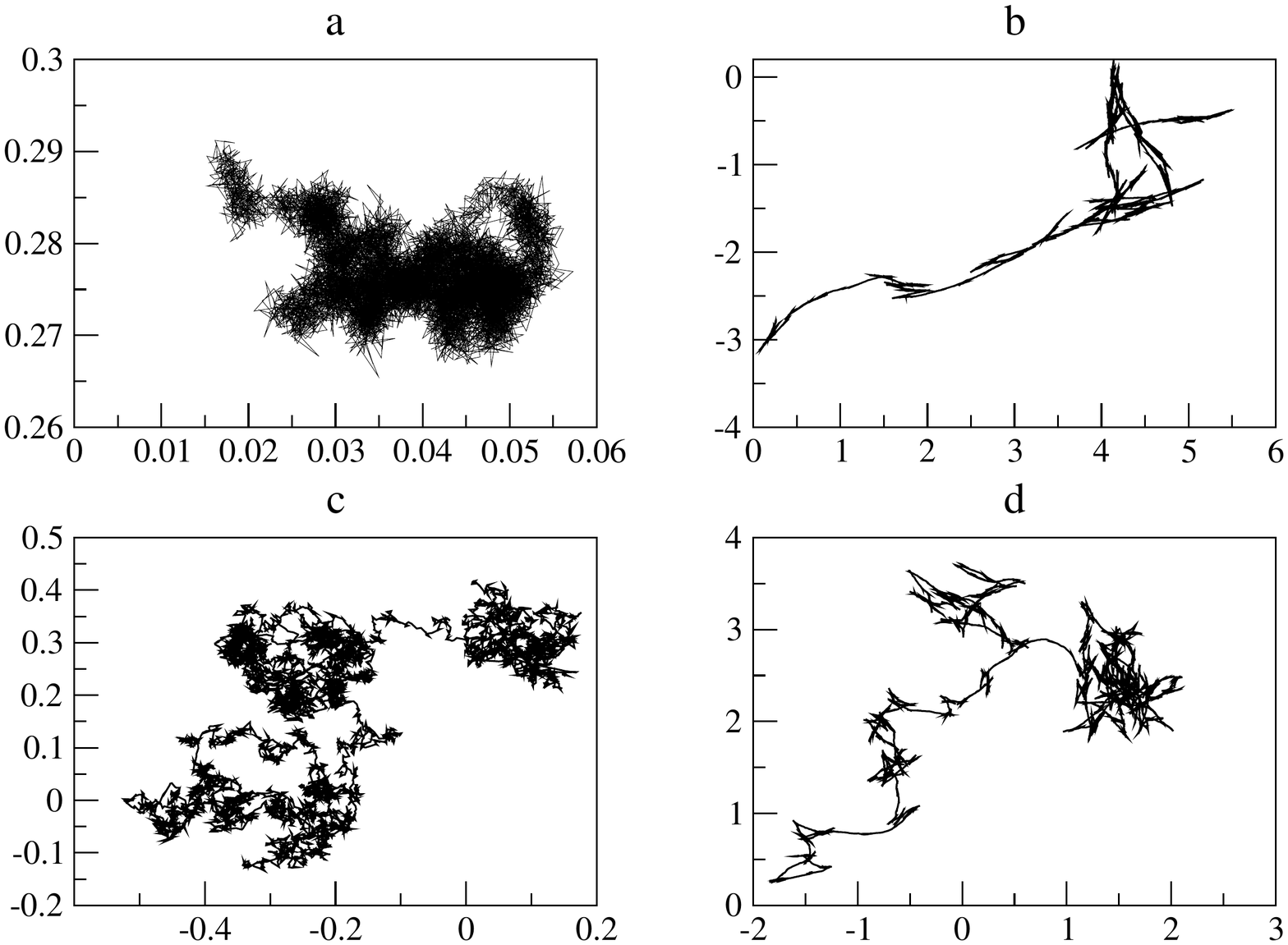}
\caption[Figure 2]{a and b represent the trajectory of the CM of the model system without and with volume exclusion respectively for parameters $\alpha=1.0$, $\gamma=0.1$. c and d are the same trajectories with volume exclusion on at the same $\alpha$ as in a and b and $\gamma = 0.01$ and $0.05$ respectively.}
\end{center}
\end{figure}

\newpage

\begin{figure}
\begin{center}
\subfigure
{\label {a}\includegraphics[width=12cm,angle=0]{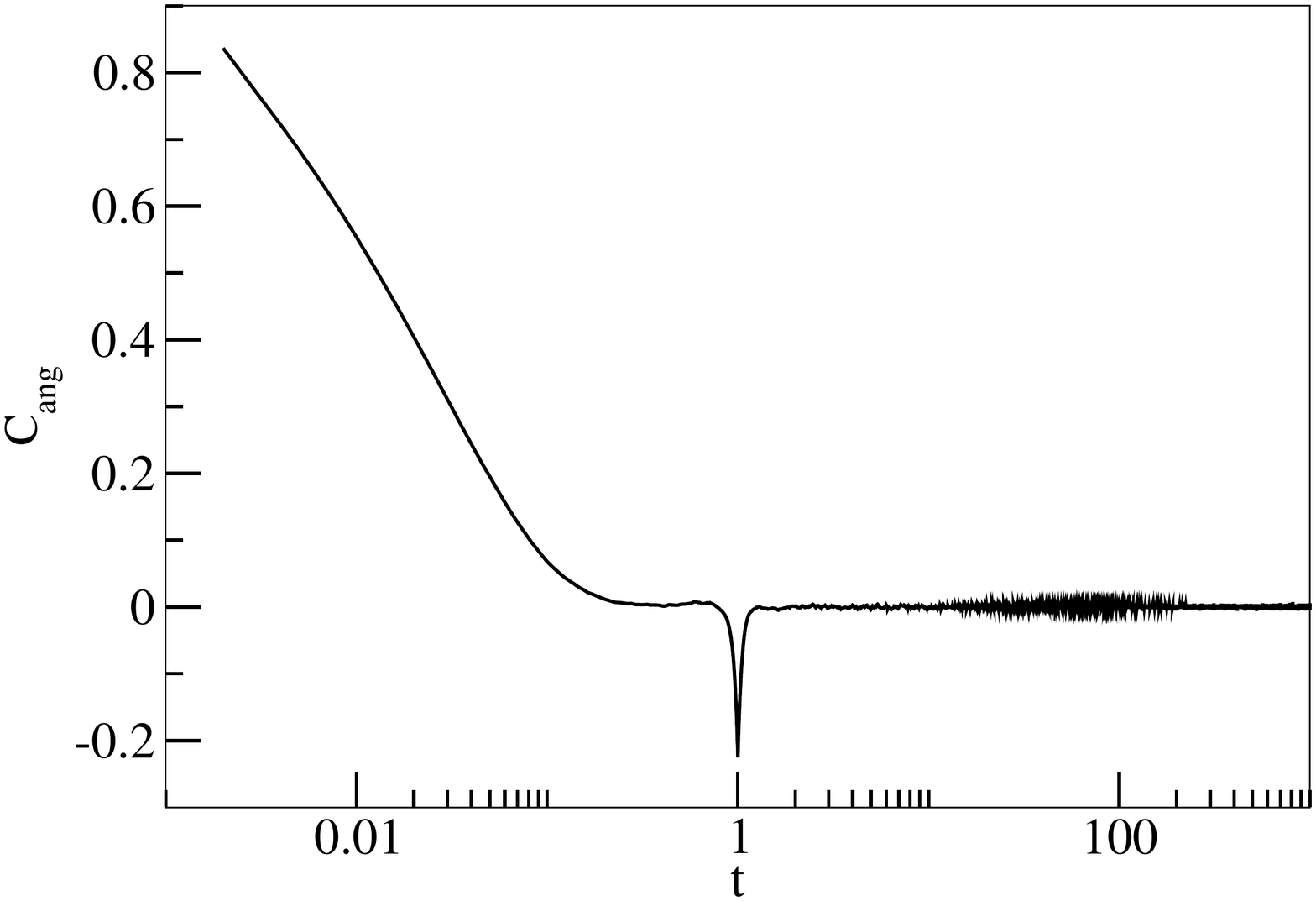}}
\subfigure
{\label {b}\includegraphics[width=12cm,angle=0]{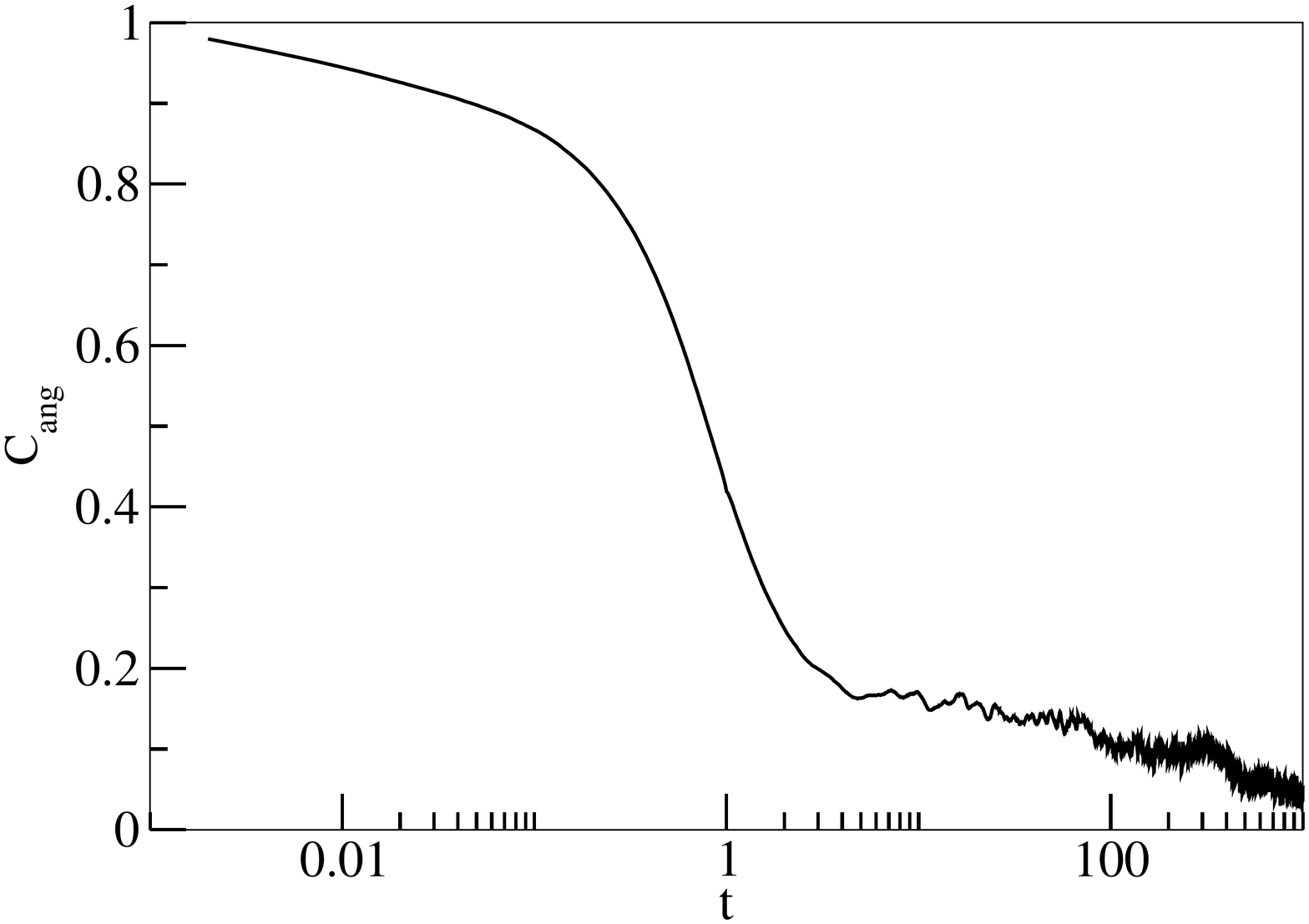}}
\end{center}
\caption [Figure 3]{Angular correlations against time without (b) and with (a) volume exclusion.}
\end{figure}

\newpage

\begin{figure}
\begin{center}
\includegraphics[width=12cm,angle=0]{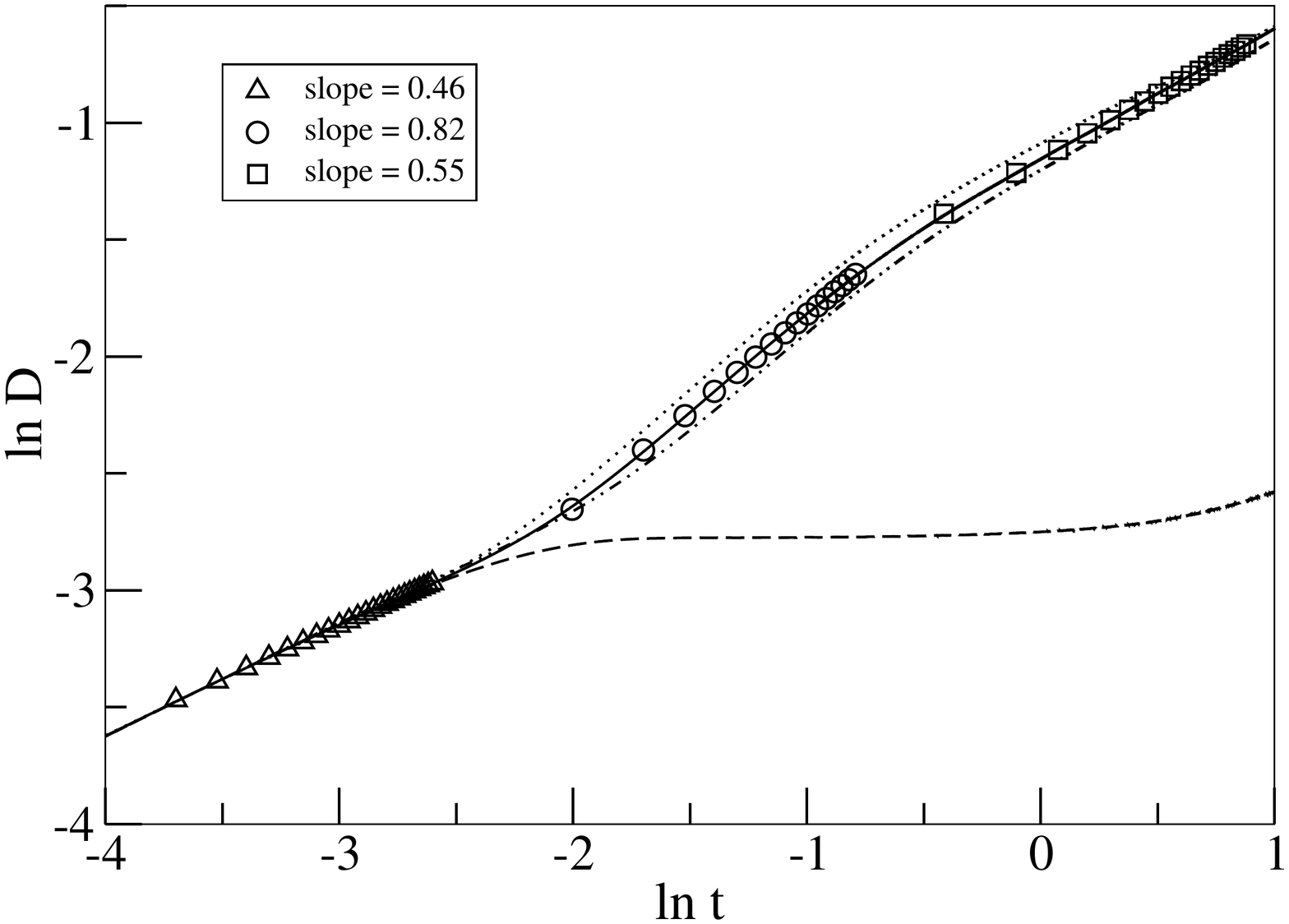}
\caption[Figure 4]{Plot of the rms deviation from the initial position of the CN to that at different times. The lowermost (dashed) curve is the case when no volume exclusion is there ($\alpha =1.0$). The upper three are with volume exclusion where $\alpha = 1.0$ (dash-dot curve), $\alpha = 2.0$ (continuous curve) and $\alpha = 3.0$ (dotted curve). The linear regression has been done on the middle (contimuous) curve and the value of the exponents are mentioned within the figure.  }
\end{center}
\end{figure}

\end{document}